\documentclass[twocolumn,trackchanges]{aastex62}
 \usepackage{ifpdf}
 \usepackage{graphicx, subfigure}
 \usepackage{natbib}
 \usepackage{times}
 \usepackage{amsmath}
 \usepackage{textcomp}
 \usepackage{lineno}
 \usepackage[colorinlistoftodos,prependcaption,textsize=tiny]{todonotes}
 \usepackage{xcolor}
 \usepackage{gensymb}

\newcommand{\Msun}{\ensuremath{M_{\odot}} }
\newcommand{\lum}{erg\,s$^{-1} $}

\graphicspath{ {./images/} }

\shortauthors{Rajagopal et al.}

\begin{document}

\title{{\it NuSTAR} Observations and multi-wavelength modeling of the high-redshift BL Lac Object 4FGL J2146.5-1344}

\email{changar@g.clemson.edu}
\author[0000-0002-8979-5254]{ M. Rajagopal}
\affil{Department of Physics and Astronomy, Clemson University, SC 29634-0978, U.S.A.\footnote{Astronomer at SARA Observatory}}
\author{L. Marcotulli}
\affil{Department of Physics and Astronomy, Clemson University, SC 29634-0978, U.S.A.}
\author{ M. Ajello}
\affil{Department of Physics and Astronomy, Clemson University, SC 29634-0978, U.S.A.}
\author{ A. Kaur}, 
\affil{Department of Astronomy and Astrophysics, 525 Davey Lab, Pennsylvania State University, University Park, 16802, U.S.A.}
\author{ V. Paliya}
\affil{Deutsches Elektronen-Synchrotron DESY, D-15738 Zeuthen, Germany}
\author{ D. Hartmann}
\affil{Department of Physics and Astronomy, Clemson University, SC 29634-0978, U.S.A. and SARA}

\begin{abstract}
High synchrotron peak (HSP; $\nu_{sy}^{pk} > 10^{15}$\,Hz) BL Lacs are some of the most extreme accelerators in the Universe. Those found at high redshifts ($z>1$) challenge our understanding of blazar evolution models and are crucial for cosmological measurements of the Extragalactic Background Light. In this paper, we study a high-$z$ BL Lac, 4FGL J2146.5-1344, detected to be at $z$=1.34 using the photometric dropout technique. We collected multi-wavelength data for this source from optical up to $\gamma$-rays, in order to study its spectral energy distribution (SED). In particular, this source was observed for the first time with {\it NuSTAR}, which accurately measures the synchrotron emission of this blazar up to 50\,keV. 
Despite being classified as an HSP BL Lac object, the modeling of the SED reveals that this source likely belongs to the ``masquerading BL Lac" class, which comprises of FSRQs appearing as disguised BL Lac objects.
\end{abstract}

\section{Introduction} \label{sec:intro}
Blazars form the largest class of active galactic nuclei (AGN) detected in the Fourth {\it Fermi}-Large Area Telescope ({\it Fermi}-LAT) source catalog (4FGL; \citealp{4fgl}), making up about 97\% of the total AGN population in the 50 MeV - 1 TeV range. Displaying highly variable non-thermal emission credited to relativistic jets aligned very close to our line of sight (\citealp{blandford1978}), blazar spectral energy distribution (SED) is typically characterised by two broad bumps, one at lower energies (Infrared to X-rays), attributed to synchrotron emission and inverse Compton scattering at higher energies (X-rays to $\gamma$-rays) \citep{maraschi1994, abdo2011}. The two sub-classes of blazars are BL Lacertae objects (BL Lacs) and Flat Spectrum Radio Quasars (FSRQs), mainly distinguished by their optical spectroscopic characteristics. BL Lacs have been observed to have either no or very weak (equivalent width \textless 5{\AA}) emission lines \citep{urry1995}, whereas FSRQs exhibit broad emission lines. The characteristic of BL Lacs indicates either an especially strong non-thermal continuum or atypically weak thermal disk/broad line emission which is mainly attributed to low accretion activity, jet dilution, or possibly both \citep{giommi2012}. Based on the frequency of synchrotron peak ($\nu_{pk}^{sy}$), blazars are further classified into 3 categories \citep{abdo2010}, namely: low-synchrotron peak blazars (LSP; $\nu_{pk}^{sy}<10^{14}$ Hz), intermediate-synchrotron peak blazars (ISP; $10^{14}$ Hz $<\nu_{pk}^{sy}<10^{15}$ Hz) and high-synchrotron peak blazars (HSP; $\nu_{pk}^{sy}>10^{15}$ Hz). A sizable population of BL Lacs lie within the ISP and HSP category \citep{ackermann2015b},  exhibiting $\nu_{pk}^{sy}$ up to $10^{17}$ Hz. BL Lacs with such large $\nu_{pk}^{sy}$ are able to accelerate electrons to beyond 100 TeV \citep{costamente2001, tavecchio2011}, making them some of the most powerful accelerators in the Universe. 

These BL Lacs are extremely crucial for the studies of extragalactic background light (EBL) \citep{ack2012, ajello2018}, which constitutes the emission of all stars and accreting compact objects in the observable universe since the re-ionization epoch. Presence of the zodiacal light and Galactic emission \citep{hauser2001} make direct studies of the EBL a challenging task. An indirect approach employed in measuring EBL intensity involves using $\gamma$-ray photons emitted by highly energetic sources (blazars). The interaction between these photons and the EBL ones causes an attenuation in the spectra \citep{stecker1992, ack2012} of these $\gamma$-ray sources through a production of electron-positron pairs. This signature allows us to constrain EBL and study its evolution with redshift \citep{Aharonian2006}. Stronger attenuation is achieved when the $\gamma$-ray source is present at higher redshifts ($z$), which leads to better EBL constraints. Therefore, on account of being bright $\gamma$-ray sources with significant emission $>$10 GeV, high redshift (high-$z$) BL Lacs represent the perfect probes in indirect studies of the EBL. 

Years of follow-up observations utilizing a range of techniques \citep{rau2012, shaw2013a, shaw2013b, ajello2013, kaur2017} has allowed us to gather redshift constraints for the $\sim$200 brightest {\it Fermi} BL Lacs. This approach yielded the discovery of a sizable number of BL Lacs at redshifts up to $\approx$2, some of which possess hard GeV spectra (photon index $<$ 2) and surprisingly belong to the HSP BL Lacs class. 

\citet{ghis2012} and \citet{padovani2012} have proposed these candidates to be ``blue FSRQs" (alternatively also called masquerading BL Lacs), i.e., sources whose relativistic jet aimed at us swamps any broad emission lines and thus are hidden by a bright synchrotron emission. Recent evidence for TXS 0506+056 (originally classified as BL Lac), the first plausible cosmic non-stellar neutrino source detected by the \citet{icecube2018}, suggests it could belong to the masquerading BL Lac class \citep{padovani2019}. Although still very uncertain, this class may harbor cosmic neutrino emitters similar to   TXS 0506+056. However, identifying such objects is challenging.

In this work, we focus on the high-$z$ source 4FGL J2146.5-1344, found by \citet{kaur2017} to be at $z=$1.34 using the photometric redshift technique.
With the goal of identifying the nature of this source, we collected data from optical, UV and X-ray facilities. Indeed, the synchrotron emission from the jet peaks in these energy bands, hence the best approach to precisely characterize their jet properties is to accurately sample these wavelengths. To this end, we have obtained as a part of an approved Cycle 4 program\footnote{proposal number: 4231, PI: M. Ajello}, data from the Nuclear Spectroscopic Telescope Array ({\it NuSTAR}, \citealp{harrison2013}). Launched in June 2012, {\it NuSTAR} has been a critical instrument for HSP BL Lac studies. Spanning an energy range from 3-79 keV, {\it NuSTAR}'s capabilities enables us to sample the falling part of the synchrotron emission, allowing us to solidly constrain the shape of the underlying electron distribution as well as the jet properties such as the bulk Lorentz factor, magnetic field strength and jet power \citep{ghis2012}.

\begin{table*}
\large
\caption{Analysis results and model parameters obtained from optical, UV and X-ray analysis}
\resizebox{\textwidth}{!}{
\begingroup
\setlength{\tabcolsep}{10pt}
\renewcommand{\arraystretch}{1.5}
\begin{tabular}{ c c c c c c }
\tableline
\tableline
\hline
\multicolumn{6}{c}{\bf SARA AB Magnitudes}\\
\tableline
\tableline
g$\arcmin$ & r$\arcmin$ & i$\arcmin$& z$\arcmin$ & &\\
\hline
17.399$\pm$0.018 & 17.179$\pm$0.014 & 17.101$\pm$0.018 & 16.686$\pm$0.039 & &\\
\tableline
\tableline
\multicolumn{6}{c}{\textbf{\textit{Swift}-UVOT  Magnitudes}} \\
\tableline
\tableline
UVW2 & UVM2 & UVW1 & U & B & V \\
\hline
18.132$\pm$0.076 & 18.002$\pm$0.095 & 17.940$\pm$0.095 & 17.683$\pm$0.092 & 17.284$\pm$0.102 & 17.120$\pm$0.170 \\
\tableline
\tableline
\multicolumn{6}{c}{\textbf{\textit{Fermi}-LAT}}\\
\tableline
\tableline
$\Gamma_{\gamma}^{a}$ & Flux$^{b}$ ($10^{-11}$\,erg cm$^{-2}$ s$^{-1}$) & Counterpart & Radio Flux (mJy) & & \\
\hline
1.71$\pm$0.04 & 1.33$\pm$0.11 & NVSS J214637-134359 & 22.951 & &\\
\tableline
\tableline
\multicolumn{6}{c}{\textbf{\textit{XMM + NuSTAR}}}\\
\tableline
\tableline
$\Gamma_{\rm X}^{c}$ & Flux$^{d}$ ($10^{-12}$\,erg cm$^{-2}$ s$^{-1}$) & $\chi^2_{\nu}$ (D.O.F.) & & &\\
\hline
2.48$\pm$0.02 & 5.6116$\pm$0.003 &  0.96 (620) & & &\\
\tableline
\tableline
\tableline
\multicolumn{6}{l}{%
  \begin{minipage}{\textwidth} 
\tablenotetext{a}{Power-law $\gamma$-ray index from 4FGL.}
\tablenotetext{b}{$\gamma$-ray flux between 1-100 GeV from 4FGL.}
\tablenotetext{c}{X-ray power-law index obtained from XSPEC analysis}
\tablenotetext{d}{Integrated X-ray flux from 0.1 to 80\,keV obtained from XSPEC analysis}
\end{minipage}%
}
\end{tabular}
\endgroup
}
\label{Tab:d1}
\end{table*}

Furthermore quasi-simultaneous optical data from the Southeastern Association for
Research in Astronomy (SARA, \citealp{keel2017}) consortium’s 0.65m telescope in Chile and UV/Optical data from Neil Gehrels $\emph{Swift}$ Observatory's UV/Optical Telescope ($\emph{Swift}$-UVOT; \citealp{gehrels2004}) were utilized along with X-ray and $\gamma$-ray data obtained from {\it XMM-Newton} \citep{jansen2001} and {\it Fermi}-LAT \citep{3fgl}, respectively, in order to construct a multi-wavelength SED of the source. 

The order of the paper is as follows: Section \ref{sec:target} describes the details about the source selection. Section \ref{sec:obs} elaborates the observations and data analysis methods and Section \ref{sec:xray} elaborates the X-ray analysis method. Section \ref{sec:model} describes  the modeling procedure while Section \ref{sec:dis} discusses the results and conclusions. We use a flat $\Lambda$CDM cosmological model with H$_{0}=67.8$ km s$^{-1}$ Mpc$^{-1}$, $\Omega_{m}=0.30$, and $\Omega_{\Lambda}=0.69$.\\

\section{TARGET SELECTION}\label{sec:target}
4FGL J2146.5-1344 (J2146 from here on) is an HSP BL Lac first detected in the 1FGL catalog \citep{1fgl} and then subsequently reported in all {\it Fermi}-LAT catalogs \citep{2fgl, 3fgl, 4fgl}. Found to be a high-$z$ source ($z=$1.34) by \citealp{kaur2017} as a part of the photometric campaign for BL Lacs begun by \citealp{rau2012}, J2146 exhibits a synchrotron peak frequency of $\sim$10$^{16}$ Hz,  a luminosity in excess of $\sim$10$^{47}$ erg s$^{-1}$, and a very hard $\gamma$-ray spectrum with photon index $\sim$1.6.  
The considerable emission at 100 GeV \citep{3fhl} also makes it a powerful tool to probe EBL, since at $z$=1.34 the universe is already opaque to the propagation of E $\sim$ 100 GeV photons (optical depth, $\tau>$1.3; \citealp{dominguez2011}). This implies that these high-$z$ HSP sources can help constrain the cosmic $\gamma$-ray horizon, which is defined by the energy at which the optical depth ($\tau$) is 1, as a function of redshift.

\section{OBSERVATIONS AND DATA ANALYSIS}\label{sec:obs}

\subsection{{\it Fermi}-LAT}
Data at $\gamma$-ray energies for J2146 is provided in the fourth {\it Fermi}-LAT source catalog \citep{4fgl}, which includes all the sources detected at energies between 50 MeV  and 1 TeV. $\gamma$-ray flux from 1-100 GeV and the uncertainty associated is shown in Table~\ref{Tab:d1}. Both 3FGL and 4FGL catalogs report the source as non-variable and since the source is absent from the 2FAV catalog \citep{Abdollahi2017}, we are able to use the data from 4FGL catalog for the SED construction of this source. 




\subsection{{\it NuSTAR}}
J2146 was observed by {\it NuSTAR} on 18 May 2018 for 42.3 ks. Data was processed for both the instrument Focal Plane Modules A (FPMA) and B (FPMB) using the  {\it NuSTAR} Data Analysis Software, {\it NuSTARDAS}, integrated in the HEASoft v.6.21\footnote{\url{https://heasarc.nasa.gov/lheasoft/}} software package. Calibrations were performed using the {\tt\string nupipeline} task (response file obtained from the latest {\tt\string CALDB} database, v.20180419). For the source extraction, a circular region of radius 30$\arcsec$ was centered on the coordinates of the target and a background circular region of $\sim$30$\arcsec$ radius was placed in the same frame, avoiding possible contamination from source photons to provide a good signal-to-noise ratio. These events were operated on by the task {\tt\string nuproducts} in order to generate the spectra and matrix files. The spectra obtained were rebinned to have 15 counts per bin.

\subsection{{\it XMM-Newton}}
The X-ray Multi-Mirror Mission spacecraft ({\it XMM-Newton}), launched by the European Space Agency (ESA), consists of three X-ray telescopes equipped at their foci with set of three X-ray CCD cameras: MOS1, MOS2 (Metal Oxide Semi-conductor), and the European Photon Imaging Camera (EPIC) pn CCD. In the energy range from 0.2 to 12 keV,  {\it XMM-Newton} allows us to observe sources with extreme sensitivities over the telescope's field of view. J2146 was observed by {\it XMM-Newton} on 18 May 2018 for 17 ks. Observations for the source were obtained and processed for all three CCD arrays. The {\it XMM-Newton} Science Analysis Software (SAS) v16.0.0. was employed for data reduction and tasks {\tt\string emproc} and {\tt\string epproc} were used for generating event files for MOS and pn respectively. Source and background spectra were generated using {\tt\string evselect} after extracting source and background regions of 10$\arcsec$ and 20$\arcsec$ respectively. The spectra obtained were rebinned to have 15 counts per bin.

\subsection{{\it Swift} and SARA}
The X-ray data was supplemented by optical observations from the SARA consortium's 0.65m telescope at Cerro Tololo, Chile (SARA-CT) and UV/Optical observations from the {\it Swift} satellite \citep{gehrels2004}. 
Data was gathered with SARA-CT on 27 August 2018 sequentially in 4 filters ({\it g$'$, r$'$, i$'$, z$'$}) and {\it Swift}-UVOT (The Ultraviolet and Optical Telescope; \citealp{roming2005}) conducted observations of the source in 6 filters ({\it uvw2, uvm2, uvw1, u, b, v}) on 23 June 2018, obtaining data for $\sim$2000 s.
Optical data was reduced using the photometry technique with the help of the software package, IRAF (v2.16; \citealp{Tody1986}). Standard star calibrations were performed using the SDSS Data Release 13 \citep{sdssdr13} and Galactic foreground extinction corrections were made using \citealp{schlafly2011}.\\
{\it Swift}-UVOT data reduction employed the use of the standard UVOT pipeline procedure \citep{poole2007} to obtain the magnitudes of the source in each filter. This was achieved using HEASoft v.6.21\footnote{\url{https://heasarc.nasa.gov/lheasoft/}} software and the tasks therein. The task {\tt\string uvotimsum} was used to combine image snapshots obtained from multiple observations and task {\tt\string uvotsource} was used to obtain the magnitudes and errors. A source region of radius 5$\arcsec$ and a background region of 25$\arcsec$ was selected in order to maximize the signal-to-noise ratio and subtract the background. The obtained magnitudes were corrected for Galactic foreground extinction using Table 5 presented in \citet{kataoka2008} and converted to AB system. 
\begin{figure*}[ht!]
    \centering
	\makebox[\textwidth]{\includegraphics[width=0.6\paperwidth]{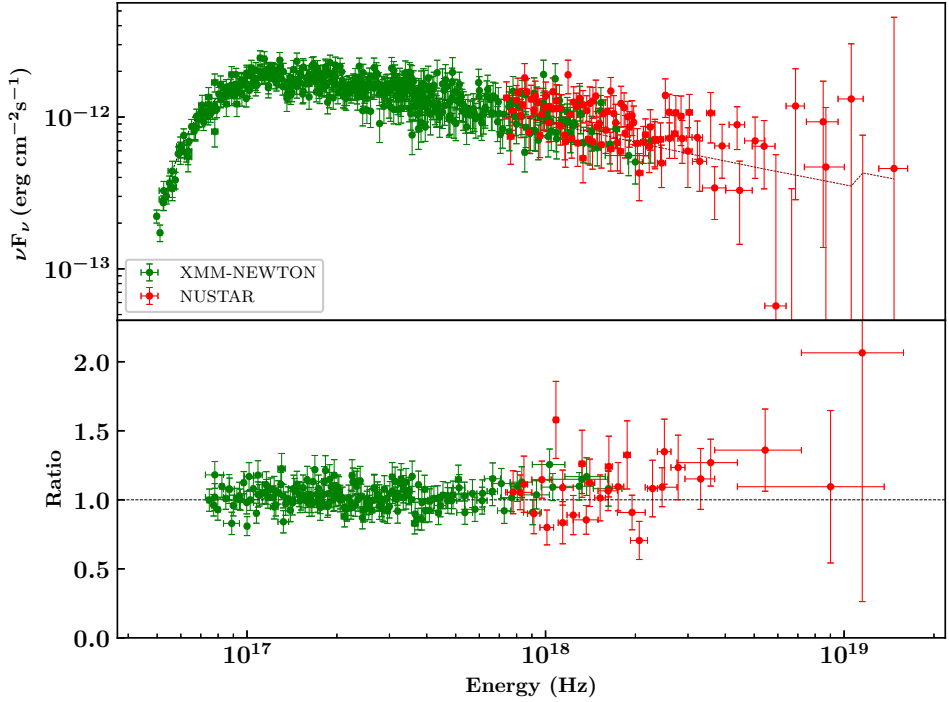}}	
		\caption{\label{fig:nustar} Combined {\it XMM-Newton} and {\it NuSTAR} spectrum. The fit was obtained using XSPEC and a simple power law model.
		}
\end{figure*}

\section{X-ray Spectral Analysis}\label{sec:xray}
The X-ray spectral analysis for {\it XMM-Newton} and {\it NuSTAR} spectra was carried out using the {\tt\string XSPEC} tool (also provided in the HEASoft package). We obtained the Galactic column density (N$_{H}$) for the source using \citet{kalberla05} and employed a $\chi^{2}$-fitting procedure which used a simple power law model with a multiplicative constant factor to fit the source. The constant was fixed at unity for EPIC pn and left free for all other instruments in order to calibrate them against each other. The flux of FPMA was found to be $\sim$20\% of FPMB. Upon investigation, it was established that the source falls in a chip gap between the two focal planes accounting for the loss of photons. In order to account for this discrepancy, we separately fitted with {\tt\string XSPEC} both {\it NuSTAR} focal modules using a simple power law. The resulting fluxes and indices are consistent for both FPMA and FPMB considering the errors so the gap does not influence our X-ray analysis. Furthermore, we tested for a possible curvature in the spectrum of the source. We jointly fitted {\it XMM-Newton} and {\it NuSTAR} with a log-parabolic model (\texttt{logpar} in {\tt\string XSPEC}) and a broken power-law one (\texttt{bknpo} in {\tt\string XSPEC}), always keeping the galactic N$_H$ fixed. The outcomes of the spectral fits did not provide any significant improvement with respect to the simple power law. Results using the F-test returned p-values $\sim10^{-1}$, hence, any curvature in the X-ray spectrum of our source can be excluded.
The parameters obtained from the model fit are shown in Table~\ref{Tab:d1} and the model fit is depicted in the Figure~\ref{fig:nustar}.

\section{modeling}\label{sec:model}
To explain the SED of J2146, a single-zone leptonic emission model is adopted. In this Section, we highlight its general outline \citep[see][ for more details]{2009MNRAS.397..985G}.
The radiation is assumed to be produced by relativistic electrons enclosed in a spherical region distant $R_{\rm diss}$ from the black hole. This region encompasses the total jet cross-section and moves with a bulk Lorentz factor, $\Gamma$. The electrons are distributed in energy according to a broken power-law shape of the type:
 \begin{equation}
 N(\gamma)  \, \propto \, { (\gamma_{\rm break})^{-p} \over
(\gamma/\gamma_{\rm break})^{p} + (\gamma/\gamma_{\rm break})^{q}}. \label{eq:shape}
\end{equation}

where $\gamma_{\rm break}$ is the energy break and {\it p} and {\it q} are the slopes before and after the break.

The particles are embedded in an uniform and randomly oriented magnetic field ($B$). As a consequence, they accelerate and thereafter radiate via synchrotron process. In presence of an external radiation field they also lose energy via Inverse Compton process. In the model, both synchrotron self Compton (SSC) and External Compton (EC) are taken into account. In the SSC case, photons produced by synchrotron emission are up-scattered to higher energies by the same electron population. For the EC, the electrons instead interact with photons external to the jet, up-scattering them to high-energies. The following are considered as reservoirs of low-energy photons:

\begin{itemize}
\item The accretion disk. It is modeled as a standard \citet{1973A&A....24..337S} disk and its spectral energy distribution is reproduced by a multicolor black-body \citep{2002apa..book.....F};
\item The X-ray corona above the disk. Its spectrum is considered to be a power-law with exponential cut-off reprocessing 30\% of disk emission;
\item The broad line region (BLR) clouds. Modeled as a spherical shell at the distance, $R_{\rm BLR} = 10^{17} L^{1/2}_{\rm disk,45}$ cm, where $L_{\rm disk,45}$ is the accretion disk luminosity in units of 10$^{45}$ \lum, from the black hole, it reprocesses 10\% of the disk emission. Its spectrum is a black-body peaking at the Lyman-$\alpha$ frequency;
\item The infrared torus. Similarly to the BLR, it is considered a spherical shell at $R_{\rm TORUS} = 10^{18} L^{1/2}_{\rm disk,45}$ cm, re-radiating 50\% of the disk emission. Its black-body spectrum peaks at the typical torus temperature of 300\,K.
\end{itemize}

The energy densities of all components depend on $R_{\rm diss}$ and are evaluated by the model.

\begin{figure*}[ht!]
    \centering
	\makebox[\textwidth]{\includegraphics[width=0.7\paperwidth]{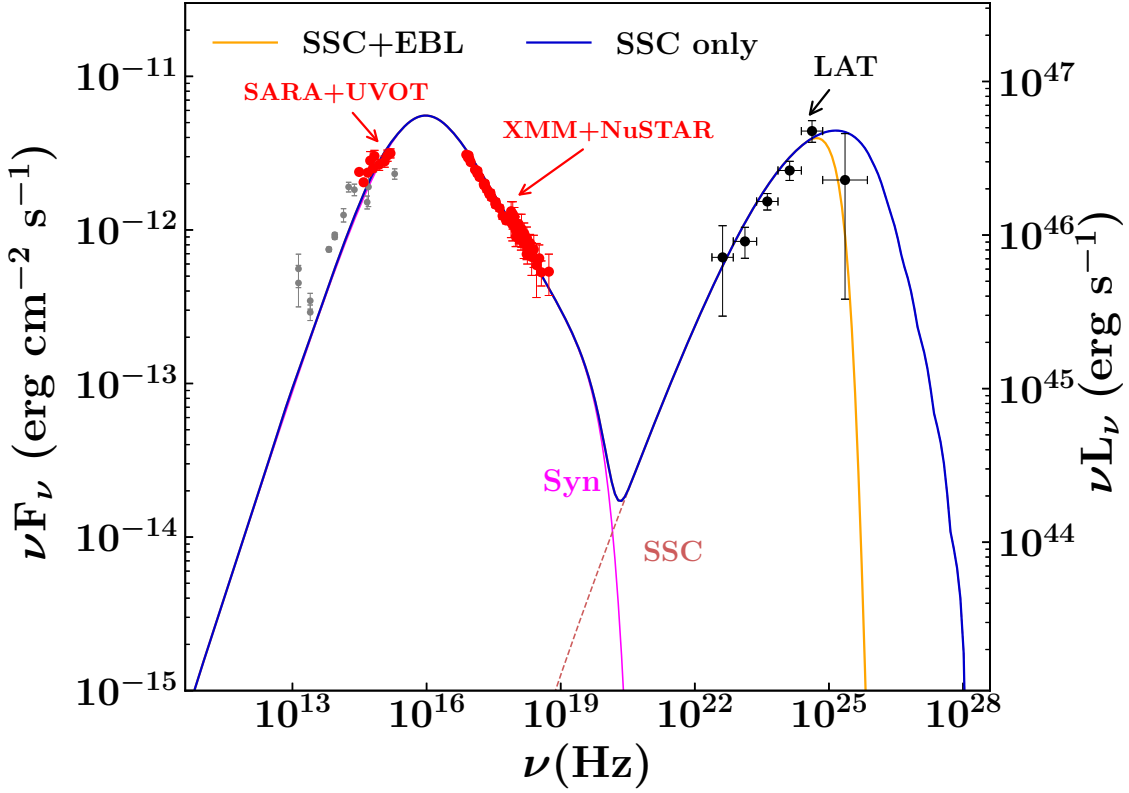}}
		\caption{\label{fig:ssc}Broadband SED of J2146 using quasi-simultaneous SARA, {\it Swift}-UVOT, {\it XMM,  NuSTAR}, and {\it Fermi}-LAT data, modeled using the one-zone leptonic emission SSC model described in the text.The grey circles represent the archival data, red circles are the points obtained from SARA+UVOT and XMM+NuSTAR data while the black points are extracted from the fourth {\it Fermi}-LAT catalog. The absorption due to the EBL is taken into account in the orange line.}
\end{figure*}

\begin{figure*}[ht!]
    \centering
	\makebox[\textwidth]{\includegraphics[width=0.7\paperwidth]{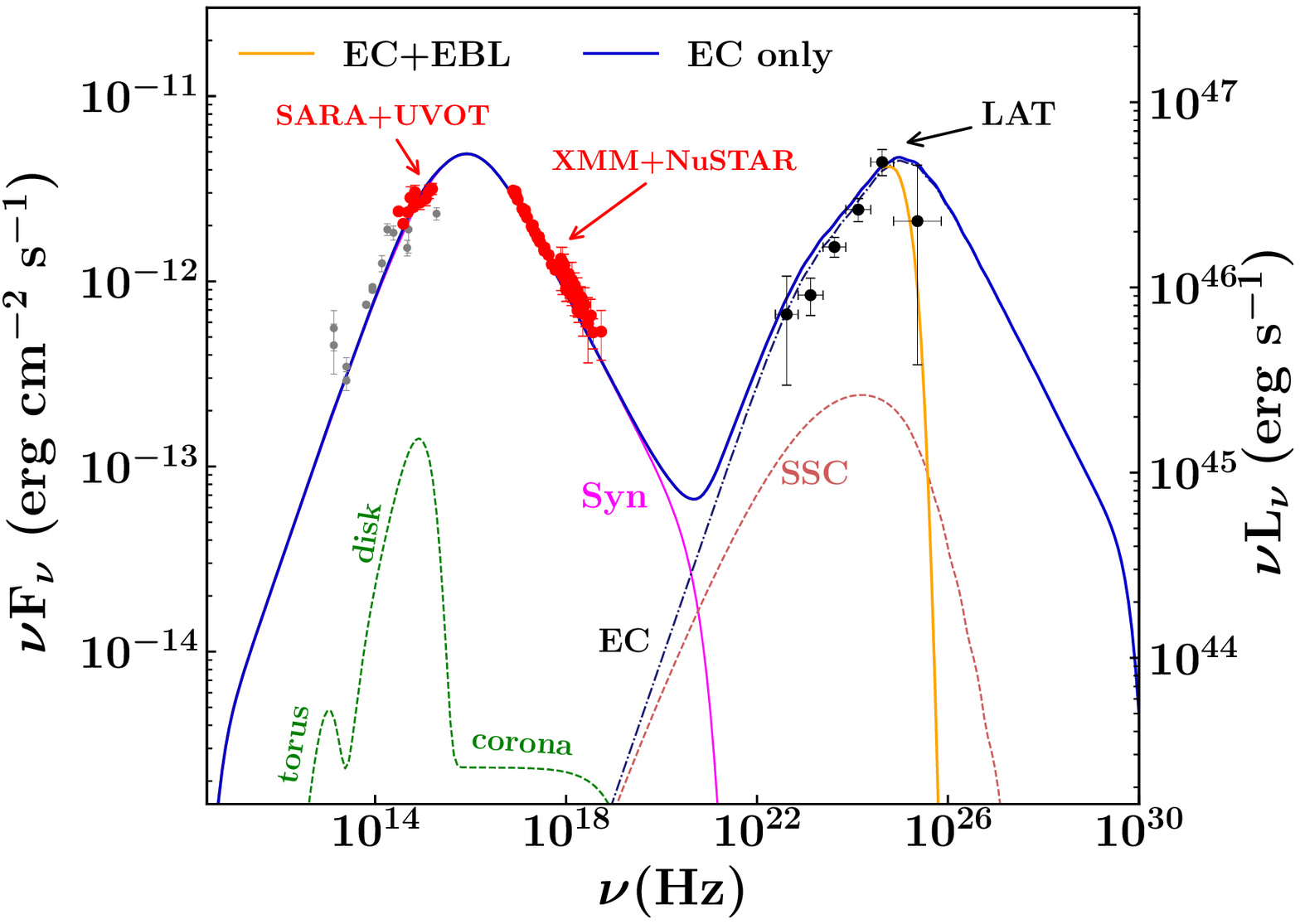}}
	    \caption{\label{fig:ec}SED modeling of J2146 using the EC model and the data points as described in the previous figure. The absorption due to the EBL is taken into account in the orange line.}
\end{figure*}

As for the black hole mass, we found that the source optical spectrum is reported in \citet{shaw2013b}, along with a redshift lower limit of 0.71 and an upper limit of 1.64, consistent with its measured photometric one \citep{kaur2017}. Although no bright emission lines are detected, the authors estimate the black hole mass of the studied BL Lacs from the $M-L$ relationship with the host galaxy, and found an average of $\sim5.6\times10^8\Msun$ (with large dispersion) \footnote{In the same work, authors report masses found spectroscopically for {\it Fermi} FSRQs which have $\overline{M}\sim1.3\times10^8\Msun$}. Similarly, \citet{sbarrato12} found an average mass for LAT detected FSRQs of $5\times10^8$\Msun and \citealp{paliya_cgrabs} derived an average mass of $8\times10^8$\Msun for radio-loud (i.e.\ jetted) CGrabs \citep{2008ApJS..175...97H} quasars using a model dependent approach. We therefore use these ranges for our assumptions on the black hole mass of J2146.   
Moreover, since the disk emission is overwhelmed by the non-thermal synchrotron one, if detected, it would result in a visible hump in the optical part of the SED. Further constraints on the disk luminosity come from empirical relations \citep[see][]{sbarrato12,ghis2012}. Following \citet{sbarrato12}, we can roughly compute the BLR luminosity ($L_{\rm BLR}$) from the $\gamma$-ray luminosity of our source ($L_{\gamma}$) through the following relationship: $L_{\rm BLR}\sim4L_{\gamma}^{0.93}$. In the case of J2146, $L_{\gamma}\sim5\times10^{46}$\,erg s$^{-1}$, hence $L_{\rm BLR}\sim10^{44}$\,erg s$^{-1}$. Assuming that the BLR reprocesses 10\% of the disk emission, this implies $L_{\rm disk}\sim10^{45}$\,erg s$^{-1}$.

The total jet power is computed as the sum of its four components: electron, proton, radiative and magnetic power. Protons are assumed to be the main carrier of kinetic jet power. They are considered to be cold, hence not radiating and only contributing to the inertia of the jet. Number densities of protons and electrons are assumed equal \citep[see][]{2008MNRAS.385..283C} and contribution of pairs is not included in the model.
Since the source has significant emission above $10$\,GeV and up to $\sim$100\,GeV (HEP = 90.02 GeV; \cite{3fhl}), knowledge of its redshift allows us to include the EBL attenuation \citep[see][]{2010ApJ...712..238F,dominguez2011,ajello2018,2019ApJ...874L...7D}. In order to do so, a multiplicative factor is introduced in the IC spectra (both SSC and EC):
\begin{equation}
I_{\rm IC,obs}=I_{\rm IC,emi}e^{-\tau(E,z)}
\end{equation}
where $I_{\rm IC,obs}$ is the observed IC intensity, $I_{\rm IC, emi}$ is the intrinsic emitted one from the source, $z$ is the redshift of the source and $\tau$ is the optical depth as function of $z$ and energy, $E$. For this work, we use $\tau$ as provided by \citet{2010ApJ...712..238F}.

With the goal of testing whether the source is more likely to be a high-luminosity BL Lac or a `masquerading BL Lac' (or a `blue FSRQ'), we separately model the source with a simple synchrotron and SSC scenario (which usually explains BL Lacs SEDs, hereafter SSC) and synchrotron, SSC and EC scenario (which usually explains FSRQs SEDs). Once the best-fit parameters are found for these two cases, we include the EBL contribution. The results of the modeling are shown in Figure~\ref{fig:ssc} and \ref{fig:ec} and are discussed in details in the Section below. All derived parameters are listed in Table~\ref{Tab:sed_par}. 

\begin{table*}[ht!]
\begin{center}
\caption{Table of used/derived parameters from the SED of J2146 for the SSC and EC model. A viewing angle of 0.5$^{\circ}$ is adopted.}
\hspace{-1.5cm}
\begin{tabular}{lccc}
\hline
Parameter & SSC  &  EC \\
\tableline
\tableline
Black hole mass ($M_{\rm BH}$) in log scale [\Msun]    & 8.7  & 8.7  \\
Accretion disk luminosity ($L_{\rm disk}$) in log scale [erg s$^{-1}$]   & 45.3  & 45.3 \\
Accretion disk luminosity in Eddington units ($L_{\rm disk}/L_{\rm Edd}$)   & 0.03  & 0.03 \\
Size of the BLR ($R_{\rm BLR}$) [pc ($R_{\rm Sch}$)] & --  & 0.04 (957.78)  \\
Dissipation distance ($R_{\rm diss}$) [pc ($R_{\rm Sch}$)] & 0.09 (2000)  & 0.06 (1325)  \\
Slope of the particle distribution below the break energy ($p$)   & 1.45  & 1.75 \\
Slope of the particle distribution above the break energy ($q$)   & 4.0  & 4.0 \\
Magnetic field ($B$) [G]                                     & 0.20 & 2.6  \\
Particle energy density ($U_{e}$) [erg cm$^{-3}$]                & $2\times{10^{-3}}$  & $5\times{10^{-4}}$ \\
Bulk Lorentz factor ($\Gamma$)                                    & 17  & 15   \\
Minimum Lorentz factor ($\gamma_{\rm min}$)                       & 1    & 1    \\
Break Lorentz factor ($\gamma_{\rm break}$)                       & $2.50\times{10^4}$    & $7.38\times10^3$ \\
Maximum Lorentz factor ($\gamma_{\rm max}$)                       & $2\times10^6$ & $2\times10^6$   \\ 
\hline
\hline
Jet power in electrons ($P_{\rm e}$) in log scale  [erg s$^{-1}$]    & 43.77  & 42.60 \\
Jet power in magnetic field ($P_{\rm B}$) in log scale [erg s$^{-1}$]   & 43.57  & 45.37 \\
Radiative jet power ($P_{\rm r}$) in log scale [erg s$^{-1}$]       & 44.93  & 44.89  \\
Jet power in protons ($P_{\rm p}$) in log scale [erg s$^{-1}$]      & 44.66  & 44.46 \\
Total jet power ($P_{\rm TOT}$) in log scale [erg s$^{-1}$]         & 44.75  & 45.39   \\
\hline
\end{tabular}
\label{Tab:sed_par}
\end{center}
\end{table*}

\begin{figure*}[ht!]
    \centering
	\makebox[\textwidth]{\includegraphics[width=0.55\paperwidth]{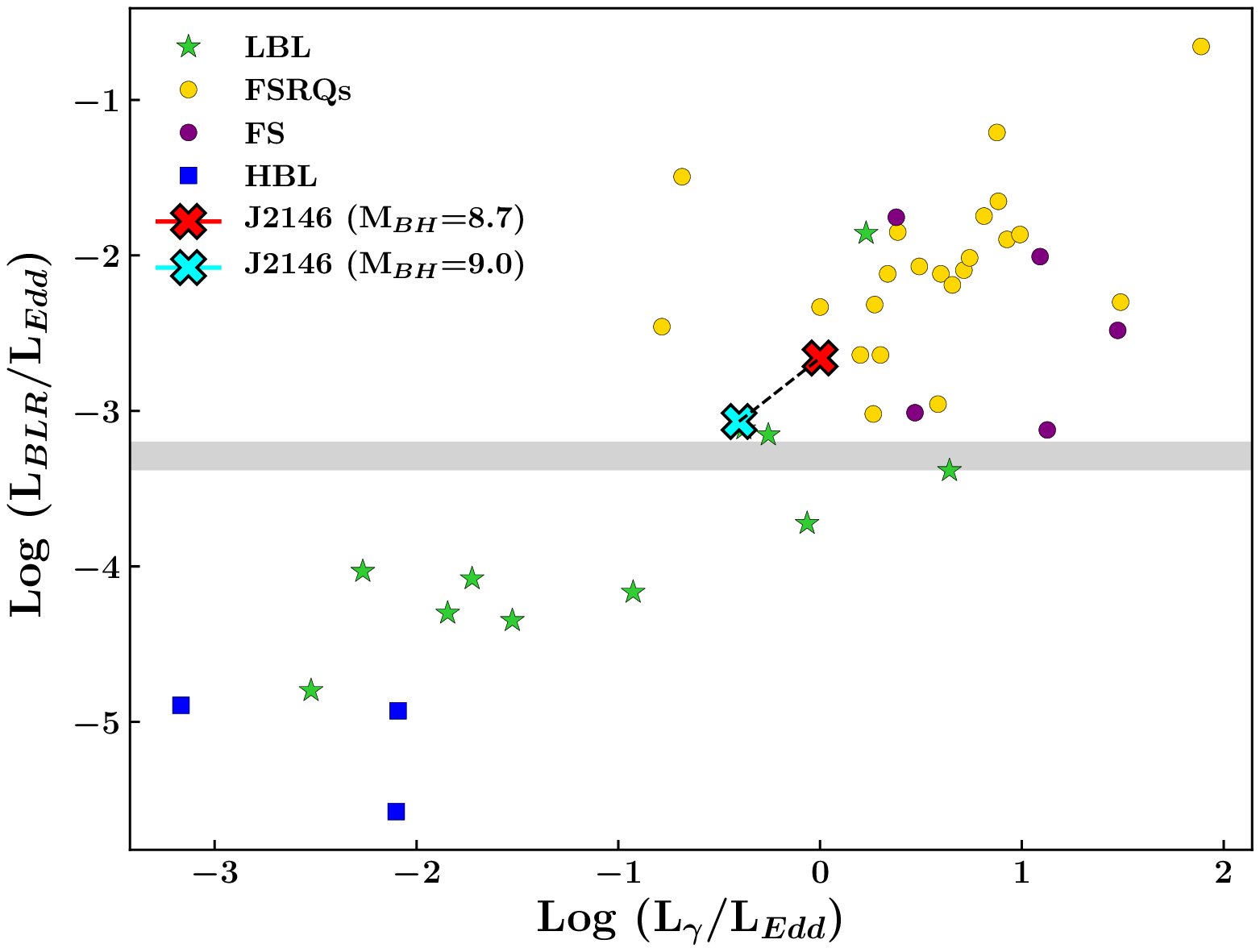}}
		\caption{\label{fig:ghis} Adapted from Figure 3 of \citet{ghis2011}, this plot shows the BLR luminosity (in units of the Eddington luminosity, $L_{\rm Edd}$) as function of the $\gamma$-ray luminosity (in units of $L_{\rm Edd}$). In their work, \citet{ghis2011} studied {\it Fermi} blazars present in the 1LAC catalog and divided them according to their SED classification. The FSRQs are represented by yellow filled circles while HBL and LBL are the green stars and blue squares, respectively. The FS sources in purple circles are the BL Lacs reclassified as FSRQs in the paper. The gray line represents the divide between BL Lacs and FSRQs. Our source, J2146, is marked with a red cross when the BH mass is the model value of 5$\times$ 10$^{8}$ \Msun. On varying the BH mass to 10$^{9}$ \Msun, we see that the source shifts to the position represented by the cyan cross. It can be seen that it still lies within the region occupied mainly by FSRQ/FS sources despite the uncertainty in the BH mass.}
\end{figure*}

\section{Discussions \& Conclusions}\label{sec:dis}
The source J2146 is one of the soft X-ray brightest ($L_{\rm X,~0.3-10\,keV}\gtrsim5\times{10}^{46}$\,erg s$^{-1}$) and high-$z$ blazars detected by the {\it Fermi}-LAT. With $\nu_{sy}^{pk} \approx 6 \times 10^{15}$\,Hz, it is among the few high-power HSP blazars so far discovered (Although \citet{masetti2013} reclassified this source as an ISP BL Lac instead of an HSP one). Only a handful of similar objects have been found \citep[see][]{padovani2012, ghis2012} and, together with our source, they challenge our understanding of the blazar population and the physical processes powering them. In fact, according to the so-called `blazar sequence' \citep[see e.g.][]{fossati1998, ghis2008, ghis2017} these kind of high-power HSP blazars should not exist. In the works of \citet{padovani2012} and \citet{ghis2012}, the authors have established that these blazars are more likely `blue-FSRQs' (or `masquerading BL Lacs', i.e. FSRQs with emission lines saturated by the non-thermal synchrotron emission) rather than BL Lac-type sources. Indeed, their high radio power ($P_{1.4\rm\,GHz}>10^{27}\rm\,W Hz^{-1}$), high synchrotron peak luminosity ($L_{\rm syn,peak}>10^{46}$\,erg s$^{-1}$) and $\nu_{sy}^{pk}>3 \times 10^{15}$\,Hz are all factors that make them resemble more FSRQs than BL Lacs \citep[see][]{giommi2012,giommi2012b}. Unveiling the nature of J2146 (i.e.\ whether it is a `masquerading BL Lac' or an HSP BL Lac) is important in the context of the blazar sequence and to test the cosmological models of the EBL. 
Moreover, J2146 is among the most luminous accelerators in our Universe and its emission up to 100\,GeV makes it an excellent probe to test cosmological models of the EBL.  

Lacking absorption lines in the optical spectra \citep[][]{shaw2013b}, 
in order to understand the nature of this blazar and to constrain the jet properties, it is necessary to obtain a multi-wavelength coverage of the source, from radio up to $\gamma$-rays.
Availability of quasi-simultaneous optical and X-ray data enables us to accurately constrain the position of the $\nu_{sy}^{pk}$ as well as shape of the underlying electron distribution, which in turn provides us with good estimates for the jet power. 
In particular, the capabilities of {\it NuSTAR} allow us to sample the falling part of the synchrotron spectrum up to 50\,keV. The source shows a very soft spectral index in this regime ($\Gamma_{\rm\,X}=2.48\pm0.02$), which is reflective of the shape of the underlying electron emitting population. Moreover, if present, a curvature in the hard X-ray spectra would have hinted towards an intrinsic curvature in the particle spectrum and would have been reflected in the falling part of the $\gamma$-ray spectrum. The lack of such feature in the X-ray continuum (see Section~\ref{sec:xray}) indicates that any curvature above $10$\,GeV is likely due to EBL absorption.\footnote{Our model does not include the Klein-Nishina effect which would produce a steeper high-energy IC spectrum \citep{2001ICRC....7.2705G,2010ApJ...721.1383A,2012JPhCS.355a2010D}. However, this effect should also be visible in the X-ray part of synchrotron spectrum, and therefore already constrained by our found $q$. The sharp cut-off in the $\gamma$-ray band could not be explained but this effect alone.}

From a modeling perspective, both the EC and SSC appear to return equally good fits.
The parameters obtained for the two models (see Table~\ref{Tab:sed_par}) and the jet power components are in agreement with what found by \citet{ghis2012}. 
We note that to explain the FSRQ-like emission (i.e.\ EC model), we need to impose the location of the emission region beyond the BLR clouds (and within the torus), conforming to what is reported in \citet{ghis2012}. Hence, the main photon energy density contributing to the EC is the torus one. This is in agreement with the recent results of  \citealp{costamante2018} and \citealp{meyer2019}, who found the emission region location outside the BLR while studying the emission of {\it Fermi}-LAT broad-line blazars. However, if this source is an FSRQ, it would need to support a very high radiative power, even larger than the kinetic one. Therefore, the jet would be radiatively and magnetically dominated in contrast with what is usually expected for FSRQs. In the SSC scenario, we point out that in order to explain the SED, we require a steeper index of the electron distribution below the break ($p$) which does not provide a very good fit to the optical and archival data. Comparing the BLR and $\gamma$-ray luminosity of J2146 with other {\it Fermi} detected blazars analyzed by \citet{ghis2011}, we can see in Figure~\ref{fig:ghis} how our source falls in the region typically occupied by FSRQs. Nonetheless, since we do not have strong constraints on the black hole mass of the object from either spectroscopy or photometric data, we tested how the position of the source would change for a higher black hole mass of $10^9$\Msun. From Figure~\ref{fig:ghis}, it can be seen how the source will still fall above the BL Lac/FSRQ divide. 
Also, since the synchrotron peak luminosity is $\sim 6\times10^{46}$\,erg s$^{-1}$ and the radio power of our source is $P_{1.4\rm\,GHz}=2.59\times10^{26}\rm\,W Hz^{-1}$ \citep{nvss}, these values are more consistent with FSRQs than BL Lacs. Overall, even though a firm conclusion on the nature of J2146 cannot be made, comparing it with the other similar objects studied in the works of \citet{padovani2012} and \citet{ghis2012}, it could likely belong to the class of `masquerading BL Lacs'. 

Finally, using the model from \citet{2010ApJ...712..238F} we are able to well model the curvature in the $\gamma$-ray spectrum. BL Lac-like sources with emission beyond $>10$\,GeV are of incredible value for studies of the EBL. Indeed, the absorption due to the annihilation of the source $\gamma$-rays with EBL photons is a tracer for the EBL intensity and it is effective beyond $>10$\,GeV. However, a lack of redshift measurements for half of the BL Lac population has hindered accurate measurements of the EBL so far. The new photometric redshift measurements of BL Lacs \citep{rau2012, kaur2017} represent a new avenue to use BL Lacs as EBL probes. As can be seen from Figure \ref{fig:ssc} and \ref{fig:ec}, the $\gamma$-ray part of J2146 SED shows a very steep fall off which is perfectly explained by EBL absorption at the redshift of the source. This points to the fact that sources such as J2146 would represent perfect probes for constraining and testing EBL models. In fact, prediction of many such models diverge as redshift increases, and the uncertainties associated get larger. Therefore, a systematic multi-wavelength study of more such blazars would allow for tighter constraints on the EBL measurements.

\acknowledgements
This work made use of data from the \textit{NuSTAR} mission, a project led by the California Institute of Technology, managed by the Jet Propulsion Laboratory, and funded by the National Aeronautics and Space Administration. We thank the \textit{NuSTAR} Operations, Software, and Calibration teams for support with the execution and analysis of these observations. This research has made use of the \textit{NuSTAR} Data Analysis Software (NuSTARDAS) jointly developed by the ASI Science Data Center (ASDC, Italy) and the California Institute of Technology (USA).\\
This research has made use of data obtained through the High Energy Astrophysics Science Archive Research Center Online Service, provided by the NASA/Goddard Space Flight Center. This research has made use of the NASA/IPAC Extragalactic Database (NED), which is operated by the Jet Propulsion Laboratory, California Institute of Technology,under contract with the National Aeronautics and Space Administration. Part of this work is based on archival data,software  or  online  services  provided  by  the ASI Data Center (ASDC).\\
The reported work is partly based on the observations obtained with the SARA Observatory telescope at Chile (SARA-CT), which is owned and operated by the Southeastern Association for Research in Astronomy (\href{www.saraobservatory.org}{saraobservatory.org}). More information about SARA can be found in DOI:   \href{https://iopscience.iop.org/article/10.1088/1538-3873/129/971/015002}{10.1088/1538-3873/129/971/015002}.\\
MR, LM and MA acknowledge funding under NASA Contract 80NSSC18K1619. The authors acknowledge the prompt observation of the source by {\it Swift}.


\bibliographystyle{apj}
\bibliography{bibliography}
\end{document}